\documentclass{aastex}
\usepackage{spr-astr-addons}
\usepackage{url}
\usepackage{amssymb}

\RequirePackage{color}

\begin{document}

\title{Tachyon Reconstruction of Ghost Dark Energy}
\slugcomment{Not to appear in Nonlearned J., 45.}
\shorttitle{Short article title}
\shortauthors{Autors et al.}

\author{A. Sheykhi\altaffilmark{}} \affil{Physics Department and Biruni Observatory, Shiraz University, Shiraz 71454,
Iran\\ Research Institute for Astronomy and Astrophysics of
Maragha (RIAAM), P. O. Box 55134-441, Maragha,
         Iran
         } \email{sheykhi@uk.ac.ir}

          \and \author{M. Sadegh Movahed \altaffilmark{}}
\affil{Department of Physics, Shahid Beheshti University, G.C.,
Evin, Tehran 19839,
         Iran
}
\email{m.s.movahed@ipm.ir} \and
\author{E. Ebrahimi} \affil{Department of Physics, Shahid Bahonar University,
P. O. Box 76175, Kerman, Iran} \email{eebrahimi@uk.ac.ir}

\begin{abstract}
Recently it has been argued that a possible source for the dark
energy may arise due to the contribution to the vacuum energy of
the QCD ghost in a time-dependent background. In this paper we
establish a connection between interacting ghost dark energy and
tachyon field. It is demonstrated that the evolution of the ghost
dark energy dominated universe can be described completely by a
single tachyon scalar field. The potential and the dynamics of the
tachyon field are reconstructed according to the evolutionary
behavior of ghost energy density
\end{abstract}



\section{Introduction}

Based on the plenty of observational evidences
\citep{riess1,riess2,Perlmut,kowalski}, in the present
time, it is accepted that the universe is undergoing a phase of
accelerated expansion due to the presence of an unknown agent
namely the ``dark energy" (DE). Identifying the origin and nature
of this unknown agent has been one of the great challenges in
modern theoretical cosmology. Many different approaches have been
proposed to solve the DE problem. These approaches can be mainly
categorized in two distinct groups. First group are the modified
gravity models which propose some serious modifications to
Einstein's theory of gravity such as $f(R)$ gravity
\citep{capoz,carroll, odinsov, movahed1,movahed2}, scalar-tensor
theories \citep{scatens, Uzan, Chiba,Bar}, Quintessence model
\citep{weterich, movahed3,movahed4,YF} and so on. The
second category are those support the idea of the existence of a
strange type of energy whose gravity is repulsive and consist an
un-clustered component through the universe. The first and
simplest candidate for DE is the cosmological constant $\Lambda$
which has constant equation of state (EoS) parameter $w=-1$
\citep{sahni}. Although this model has a good agreement with
observational data but it suffers several difficulties such as
\textit{fine tuning} and \textit{coincidence} problem
\citep{riess1,riess2}. Further observations detect a small
variation in the EoS parameter of DE in favor of a dynamic DE.
These observations show that the EoS of DE $w$ is likely to cross
the cosmological constant boundary -1 (or phantom divide), i.e.
$w$ is larger than $-1$ in the recent past and less than $-1$
today \citep{feng,alam,huterer}. The conventional scalar-field
model, the quintessence with a canonical kinetic term, can only
evolve in the region of $w > -1$, whereas the model of phantom
with negative kinetic term can always lead to $w \leq -1$. Neither
the quintessence nor the phantom alone can realize the transition
of $w$ from $w > -1$ to $w < -1$ or vice versa. A comprehensive
review on DE models can be seen in a very recent paper by M. Li,
et al. \citep{nederev}.

An interesting model of DE, called ``ghost dark energy" (GDE) was
recently proposed  \citep{urban,Otha}. The so called ``Veneziano
ghost field" is presented as a solution to $U(1)$ problem in
effective low energy $QCD$
\citep{witten,Veneziano,rosen,nath,kawar}. The ghost field seems to
be un-physical and has no contribution to the vacuum energy in the
Minkowski spacetime. However, in a dynamic background or a
spacetime with non-trivial topology the ghost field contribute to
the vacuum energy proportional to $\Lambda^3_{QCD} H$, where $H$
is the Hubble parameter and $\Lambda^3_{QCD}$ is $QCD$ mass scale
\citep{Otha}. Actually the DE models based on the ghost field
consider this vacuum energy density as a dynamical cosmological
constant to investigate its role as an alternative to resolve the
DE puzzle \citep{cai,shmov,shba,shba2,shba3}. The ghost DE model
can also categorized to the class of inhomogeneous fluid DE models
\citep{odinsov2}. One of the most important advantages of the ghost
DE model is that this model comes from the standard model of
particle physics and we do not need to introduce any new degree of
freedom.

On the other hands, the tachyon field has been proposed as a
possible candidate for DE. A rolling tachyon has an interesting
equation of state whose parameter smoothly interpolates between
$-1$ and $0$ \citep{Gib1}. Thus, tachyon can be realized as a
suitable candidate for the inflation at high energy \citep{Maz1} as
well as a source of dark energy depending on the form of the
tachyon potential \citep{Padm}. These motivate us to reconstruct
tachyon potential $V(\phi)$ from GDE model. The correspondence
between tachyon field and various dark energy models such as
holographic dark energy \citep{Setare4,NG,KK} and agegraphic dark
energy \citep{ahmad1,KK2} has been already established. The
extension has also been  done to the entropy corrected holographic
and agegraphic dark energy models \citep{ahmad2}.

The effective lagrangian for the tachyon field is given by
\citep{sen,sen2}
\begin{eqnarray}
 L=-V(\phi)\sqrt{1-g^{\mu\nu}\partial_\mu \phi \partial_\nu \phi},
 \end{eqnarray}
where $V(\phi)$ is the tachyon potential. The corresponding energy
momentum tensor for the tachyon field can be written in a perfect
fluid form
\begin{eqnarray}
 T_{\mu\nu}=(\rho_\phi+p_\phi)u_{\mu} u_\nu-p_\phi g_{\mu\nu},
 \end{eqnarray}
where $\rho_\phi$ and $p_\phi$ are the energy density and pressure
of the tachyon, respectively. The velocity $u_\mu$ is
\begin{eqnarray}
 u_\mu=\frac{\partial_\mu \phi}{\sqrt{\partial_\nu \phi \partial^\nu
 \phi}}.
  \end{eqnarray}
It was demonstrated that dark energy driven by tachyon, decays to
cold dark matter in the late accelerated universe and this
phenomenon yields a solution to cosmic coincidence problem
\citep{Sri}. Choosing different self-interaction potentials in the
tachyon field model leads different consequences for the resulted
DE model.

The rest of this paper is organized as follows. The next section
includes the relations and discussions about the reconstructed
tachyon GDE model. In section III we extend the study to the
interacting GDE model. The Summary and conclusion are given in
section IV.
\section{Tachyon ghost model }
Consider a flat Friedmann-Robertson-Walker (FRW) which its
dynamics is governed by the Friedmann equation
\begin{eqnarray}\label{Fried}
H^2=\frac{1}{3M_p^2} \left( \rho_m+\rho_D \right),
\end{eqnarray}
where $\rho_m$ and $\rho_D$ are the energy densities of
pressureless matter and GDE, respectively. The ghost energy
density is proportional to the Hubble parameter \citep{Otha}
\begin{equation}\label{GDE}
\rho_D=\alpha H.
\end{equation}
where $\alpha$ is a constant of order $\Lambda_{\rm QCD}^3$ and
$\Lambda_{\rm QCD}\sim 100 MeV$ is QCD mass scale. We define the
dimensionless density parameters as
\begin{equation}\label{Omega}
\Omega_m=\frac{\rho_m}{\rho_{\rm cr}},\ \ \
\Omega_D=\frac{\rho_D}{\rho_{\rm cr}}=\frac{\alpha}{3M_p^2 H},\ \
,
\end{equation}
where the critical energy density is $\rho_{\rm cr}={3H^2 M_p^2}$.
Using (\ref{Omega}), the Friedmann equation can be rewritten as
\begin{equation}\label{fridomega}
\Omega_m+\Omega_D=1.
\end{equation}
The conservation equations read
\begin{eqnarray}
\dot\rho_m+3H\rho_m&=&0,\label{consm}\\
\dot\rho_D+3H\rho_D(1+w_D)&=&0\label{consd}.
\end{eqnarray}
Taking the time derivative of relation (\ref{GDE}) and using the
Friedmann equation we find
\begin{equation}\label{dotrho1}
\dot{\rho}_D=\rho_D \frac{\dot{H}}{H}=-\frac{\alpha }{2 M_p^2}
\rho_D(1+u+w_D).
\end{equation}
where
\begin{equation}\label{u1}
u=\frac{\rho_m}{\rho_D}=\frac{\Omega_m}{\Omega_D}=\frac{1-\Omega_D}{\Omega_D},
\end{equation}
is the energy density ratio. Inserting relation (\ref{dotrho1}) in
continuity equation (\ref{consd}), after using (\ref{u1}) we find
\begin{equation}\label{wD1}
w_D=-\frac{1}{2-\Omega_D},
\end{equation}
At the early time where $\Omega_D\ll 1$ we have $w_D=-1/2$, while
at the late time where $\Omega_D\rightarrow 1$ the GDE mimics a
cosmological constant, namely $w_D= -1$.

\begin{figure}[t]
\begin{center}
\includegraphics[scale=0.4]{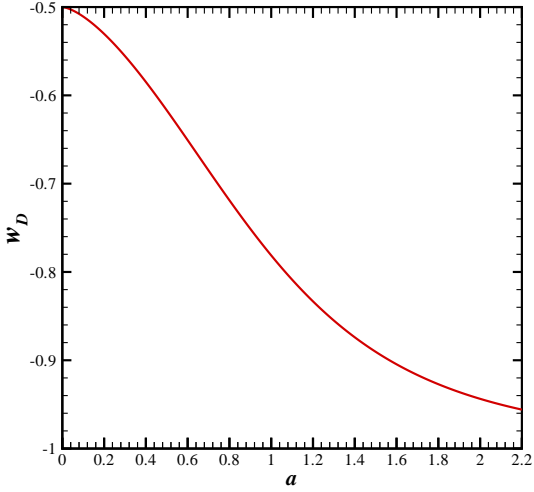}
 \caption{The evolution of $w_D$ for ghost dark
energy. Here we have taken $\Omega^{0}_{D}=0.72.$ }\label{fig1}
\end{center}
\end{figure}

\begin{figure}[t]
\begin{center}
\includegraphics[scale=0.4]{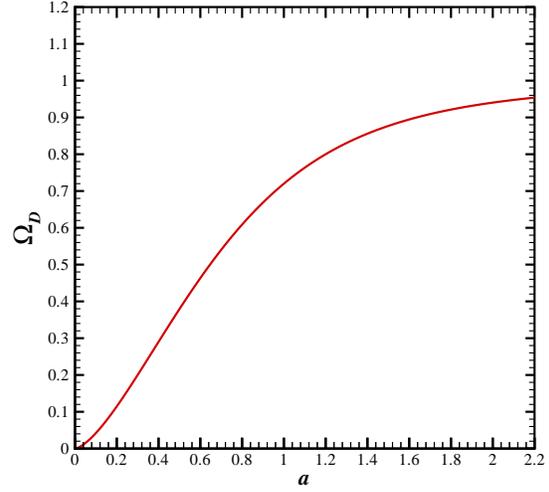}
\caption{The evolution of $\Omega_{D}$ for
ghost dark energy where again we have taken
$\Omega^{0}_{D}=0.72$.}\label{fig2}
\end{center}
\end{figure}

The equation of motion of GDE is obtained as \citep {shmov}
\begin{equation}\label{Omegaprime2}
\frac{d\Omega_D}{d\ln a}=3 \Omega_D
\frac{(1-\Omega_D)}{2-\Omega_D}.
\end{equation}
In Figs. \ref{fig1} and \ref{fig2}  we have plotted the evolution of $w_D$ and
$\Omega_D$ versus scale factor $a$. From Fig. \ref{fig1} we see that
$w_D$ of the GDE model cannot cross the phantom divide and mimics
a cosmological constant at the late time.

Next we suggest a correspondence between ghost energy density and
tachyon field. The energy density and pressure of tachyon field
are given by
\begin{equation}
\rho_\phi=-T_0^0=\frac{V(\phi)}{\sqrt{1-\dot\phi^2}},\label{rhophi1}
\end{equation}
\begin{equation}
p_\phi=T_i^i=-V(\phi)\sqrt{1-\dot\phi^2}.
\end{equation}
Thus the equation of state parameter of tachyon field is given by
\begin{equation}\label{wphi1}
w_\phi=\frac{p_\phi}{\rho_\phi}=\dot\phi^2-1.
\end{equation}
To establish the correspondence between GDE and tachyon field, we
equate  $w_D$ with $w_\phi$. From Eqs. (\ref{wD1}) and
(\ref{wphi1}) we find
\begin{equation}\label{dotphi1}
\dot\phi^2=\frac{1-\Omega_D} {2-\Omega_D}
\end{equation}
Using the second Eq.(\ref{Omega}) as well as relation
$\dot{\phi}=H \frac{d\phi}{d\ln a}$ we can rewrite the dynamics of
scalar field as
\begin{eqnarray}
\frac{d\phi}{d\ln a} &=&\frac{3M_p^2}{\alpha}\times \Omega_D
\sqrt{\frac{1-\Omega_D} {2-\Omega_D}}.\label{dotphi2}
\end{eqnarray}
Integrating yields
\begin{eqnarray}
\phi(a)-\phi(a_0)=\frac{3M_p^2}{\alpha} \int_{a_0}^{a}{ \frac
{da}{a} \Omega_D \sqrt{\frac{1-\Omega_D}
{2-\Omega_D}}},\label{phi1}
\end{eqnarray}
where $a_0=1$  is the  present value of the scale factor. To
reconstruct the tachyon potential, we identify
$\rho_{\phi}=\rho_{D}=\alpha H$ and combine Eqs.  (\ref{Omega})
and (\ref{dotphi1}) with (\ref{rhophi1}). We find
\begin{eqnarray}\label{vphi1}
V(a)&=&\frac{\alpha^2}{3 M_p^2} \times \frac
{\Omega_D^{-1}}{\sqrt{2-\Omega_D}}.
\end{eqnarray}
\begin{figure}[t]
\begin{center}
\includegraphics[scale=0.4]{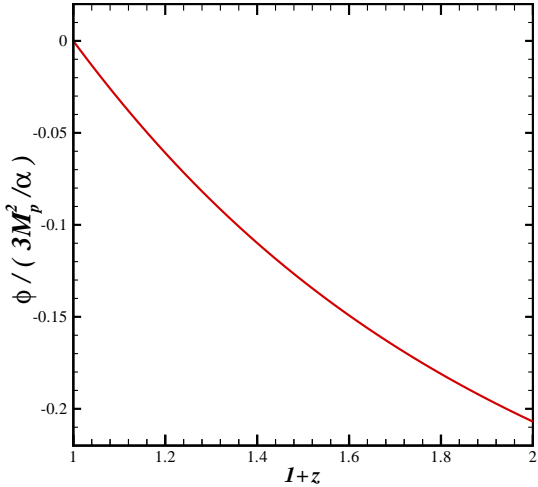}
\caption{The evolution of the scalar field
$\phi$ as a function of redshift for tachyon ghost dark energy. }\label{fig3}
\end{center}
\end{figure}
\begin{figure}[t]
\begin{center}
\includegraphics[scale=0.4]{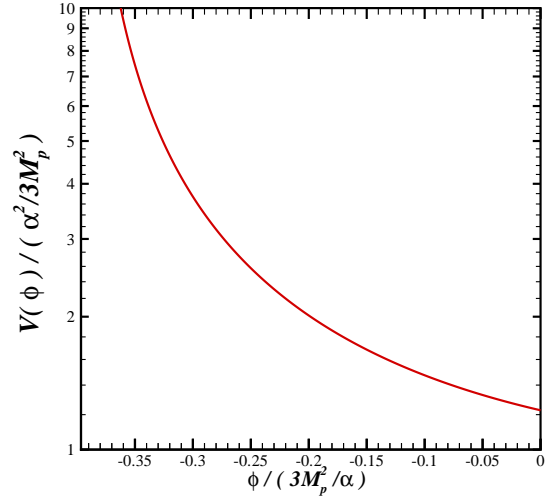}
\caption{The reconstructed potential $V(\phi)$
for tachyon ghost dark energy. }\label{fig4}
\end{center}
\end{figure}
Basically, from Eqs. (\ref{Omegaprime2}) and (\ref{phi1}) one can
derive $\phi=\phi(a)$ and then combining the result with
(\ref{vphi1}) one finds $V=V(\phi)$. Unfortunately, the analytical
form of the potential in terms of the ghost tachyon field cannot
be determined due to the complexity of the equations involved.
However, we can obtain it numerically. The evolution of the field
and the reconstructed tachyon potential $V(\phi)$  are plotted in
Figs. \ref{fig3} and \ref{fig4}, where we have taken $\phi(a_0=1)=0$ for simplicity.
From Fig. \ref{fig3} we can see the dynamics of the scalar field
explicitly. In this figure we can see that the scalar field $\phi$
increases from below to zero at the present time which is not
similar to other reconstructed models of DE. Fig. \ref{fig4}, indicates
that the reconstructed scalar potential shows a nonzero minima
which reminds the cosmological constant behavior of the model in
the present time.
\section{Interacting  tachyon ghost model}
In this section we extend our study to the interacting case. We
shall assume the two dark components namely dark matter and GDE
interact to each other thus, $\rho_{m}$ and $\rho_{D}$ do not
conserve separately and evolve according to their semi
conservation laws
\begin{eqnarray}
\dot\rho_m+3H\rho_m&=&Q,\label{consm2}\\
\dot\rho_D+3H\rho_D(1+w_D)&=&-Q\label{consd2},
\end{eqnarray}
where $Q$ represents the interaction term which can be, in
general, an arbitrary function of cosmological parameters like the
Hubble parameter and energy densities, $Q (H\rho_{m},H\rho_{D})$.
The simplest choice is $Q =3b^2 H(\rho_{m}+\rho_{D})$ with $b^2$
is a coupling constant \citep{Ame,Zim,wang1,sh09,sh10,pav1,BP}. The
positive $b^2$ is responsible for the energy transition from dark
energy to dark matter. Sometimes this constant is taken in the
range $[0, 1]$ \citep{zhang}. Note that if $b^2=0$ then it
represents the non-interacting FRW model while $b^2=1$ yields
complete transfer of energy from dark energy to dark matter.
Recently, it is reported that this interaction is observed in the
Abell cluster A586 showing a transition of dark energy into dark
matter and vice versa \citep{berto1}. Observations of cosmic
microwave background and galactic clusters show that the coupling
parameter $b^2< 0.025$, i.e. a small but positive constant of
order unity \citep{ich}, a negative coupling parameter is avoided
due to violation of thermodynamical laws. Therefore the
theoretical interacting models are phenomenologically consistent
with the observations. It should be noted that the ideal
interaction term must be motivated from the theory of quantum
gravity. In the absence of such a theory, we rely on pure
dimensional basis for choosing an interaction $Q$. Thus we take
the interaction term of the following form
\begin{equation}\label{Q}
Q =3b^2 H(\rho_m+\rho_D)=3b^2 H\rho_D(1+u).
\end{equation}
Inserting Eqs. (\ref{dotrho1}) and (\ref{Q}) in Eq. (\ref{consd2})
and using (\ref{u1}) we obtain the equation of state parameter of
interacting GDE
\begin{equation}\label{wD2}
w_D=-\frac{1}{2-\Omega_D}\left(1+\frac{2b^2}{\Omega_D}\right).
\end{equation}
In the late time where $\Omega_D\rightarrow 1$, the equation of
state parameter of interacting GDE necessary crosses the phantom
line, namely, $w_D=-(1+2b^2)<-1$ independent of the value of
coupling constant $b^2$.  At the present time with
$\Omega^{0}_D=0.72$ the phantom crossing can be achieved provided
we take $b^2>0.1$.
\begin{figure}[t]
\begin{center}
\includegraphics[scale=0.4]{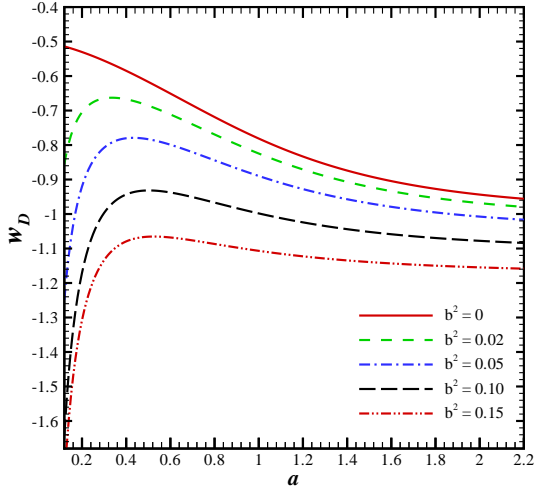}
\caption{The evolution of $w_D$ for interacting
ghost dark energy and different interacting parameter $b^2$.
}\label{fig5}
\end{center}
\end{figure}

\begin{figure}[t]
\begin{center}
\includegraphics[scale=0.4]{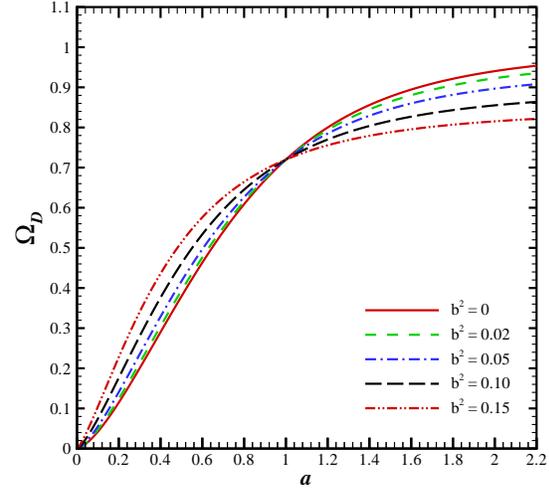}
\caption{The evolution of $\Omega_D$ for
interacting ghost dark energy. Selected curves are plotted for
different $b^2$. }\label{fig6}
\end{center}
\end{figure}
The equation of motion of interacting GDE can be obtained as
\citep{shmov}
\begin{equation}\label{Omegaprime3} \frac{d\Omega_D}{d\ln
a}=\frac{3}{2} \Omega_D\left[1-\frac{ \Omega_D}{2-
\Omega_D}\left(1+\frac{2b^2}{\Omega_D}\right)\right].
\end{equation}
The dynamics of $w_D$ and $\Omega_D$ are plotted in Figs. \ref{fig5}  and
\ref{fig6}. Selected curves are plotted for different value of the coupling
parameter $b^2$. According to figure \ref{fig6}, one finds that in
the future  $\Omega_D$ continue approaching to 1 which shows that
in this model of DE the future evolution of the universe is
determined by the dark energy component. This indicates that
probably the fate of the universe goes toward a big rip.

Having Eqs. (24)  and (25)  at hand, we are in a position to
implement a correspondence between interacting ghost energy
density and tachyon scalar field model, by comparing the ghost
density with the tachyon field model and equating the equation of
state parameter of this model with the equation of state parameter
of interacting GDE obtained in (\ref{wD2}). To this end,  we
equate  $w_D$ with $w_\phi$. From Eqs. (\ref{wD2}) and
(\ref{wphi1}) we find
\begin{equation}\label{dotphi2}
\dot\phi^2=\frac{1} {2-\Omega_D}\left(1-\Omega_D-
\frac{2b^2}{\Omega_D}\right).
\end{equation}
Using second Eq. (\ref{Omega}) as well as relation $\dot{\phi}=H
\frac{d\phi}{d\ln a}$ we can rewrite the dynamics of scalar field
as
\begin{eqnarray}
\frac{d\phi}{d\ln a} &=&\frac{3M_p^2}{\alpha}\times \Omega_D
\sqrt{\frac{1} {2-\Omega_D}\left(1-\Omega_D-
\frac{2b^2}{\Omega_D}\right)}.\label{dotphi3}
\end{eqnarray}
Integrating yields
\begin{eqnarray}
\phi(a)-\phi(a_0)=\frac{3M_p^2}{\alpha} \int_{a_0}^{a}{ \frac
{da}{a}  \sqrt{\frac{\Omega_D^2} {2-\Omega_D}\left(1-\Omega_D-
\frac{2b^2}{\Omega_D}\right)}}.\label{phi2}
\end{eqnarray}
where $a_0=1$  is the  present value of the scale factor. To
reconstruct the tachyon potential, we identify
$\rho_{\phi}=\rho_{D}=\alpha H$ and combine Eqs.  (\ref{Omega})
and (\ref{dotphi2}) with (\ref{rhophi1}). The result is
\begin{eqnarray}\label{vphi2}
V(\phi)&=&\frac{\alpha^2}{3 M_p^2}\times
\frac{1}{\Omega_D}\left(\frac{1+2b^2 \Omega_D^{-1}}{
2-\Omega_D}\right)^{1/2},
\end{eqnarray}
 The evolutionary
form of the tachyon field and the reconstructed tachyon potential
$V(\phi)$  are plotted in Figs. 7 and 8. Again we have taken
$\phi(a_0=1)=0$ for simplicity. Selected curves are plotted for
different value of the coupling parameter $b^2$. From these
figures we find out the reconstructed scalar field has a same
dynamic as the non-interacting case. For different choices of the
coupling parameter $b^2$ we find a faster rate of evolution when
$b^2$ increases. The reconstructed scalar potentials in Fig.
\ref{fig8} generally show decreasing and flattening in the near
epoches. As the non-interacting case the scalar potential has a
non-zero minimum which leads to an EoS parameter close to $-1$ for
present time and near future. If the future evolution of the
potential has a mirror image behavior of the plotted regions we
can see that increasing $b^2$ leads to steeper and steeper form of
potentials. In this form of potentials, the scalar field
oscillates around a minima and settles down in the minima. The
oscillation around the minima for $\phi^2$-like potentials (with
suitable choice of $b^2$) is completely harmonic and can acts as
pressureless fluid during the oscillation period of the scalar
file as an alternative to cold dark matter.
\begin{figure}[t]
\begin{center}
\includegraphics[scale=0.4]{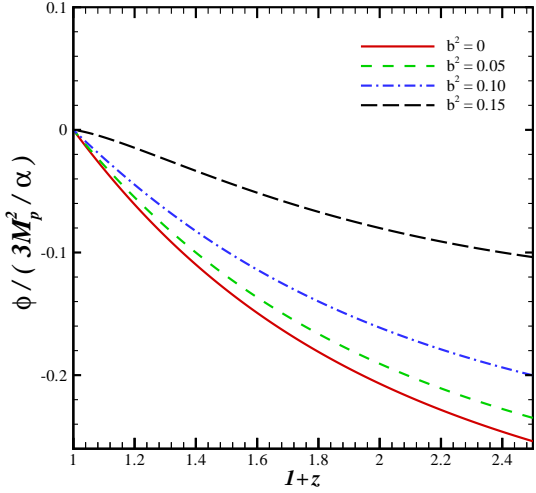}
\caption{The evolution of the scalar field
$\phi$ as a function of redshift for interacting tachyon ghost dark energy and different
coupling parameter $b^2$. }\label{fig7}
\end{center}
\end{figure}
\begin{figure}[t]
\begin{center}
\includegraphics[scale=0.4]{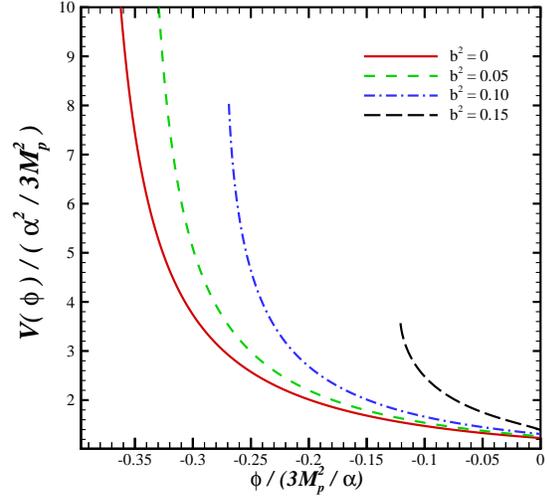}
\caption{The reconstructed potential $V(\phi)$
for interacting tachyon ghost dark energy and different coupling
parameter $b^2$. }\label{fig8}
\end{center}
\end{figure}

\section{Conclusion}\label{con}
The so called "ghost dark energy" was recently proposed to explain
the dark energy dominated universe. In a dynamic background or a
spacetime with non-trivial topology the ghost field contribute to
the vacuum energy proportional to $\Lambda^3_{QCD} H$, where $H$
is the Hubble parameter and $\Lambda^3_{QCD}$ is $QCD$ mass scale.
A suitable choice of the $H$ and $\Lambda_{QCD}$ leads to right
value of $\rho_D=\alpha H$. The advantages of this new proposal
compared to the previous dark energy models is that it totally
embedded in standard model so that one does not need to introduce
any new parameter, new degree of freedom or to modify general
relativity.

On the other hand, we know that the scalar field models of dark
energy can be considered as an effective theory of the underlying
theory of dark energy. This point motivated us to reconstruct the
tachyon model of dark energy based on the ghost energy density. To
this end, we have constructed a version of tachyon dark energy
which mimics the behavior of the ghost model of dark energy in the
early epoches and late time. Different quantities are plotted and
evolution of the model is shown in different epoches. Due to
importance of correspondence between these models (GDE and
tachyon), one can mention the cosmological constant-like behavior
of both of models in the late time. Another result of this
correspondence is approaching of the reconstructed scalar field to
zero from below which is different with respect to the other
scalar field models.


\acknowledgments{This work has been supported financially by
Research Institute for Astronomy and Astrophysics of Maragha
(RIAAM) under research project No. 1/2337.}


\end{document}